\documentclass{article}

\usepackage{arxiv}

\usepackage[utf8]{inputenc} % allow utf-8 input
\usepackage[T1]{fontenc}    % use 8-bit T1 fonts
\usepackage[hidelinks]{hyperref}       % hyperlinks
\usepackage{url}            % simple URL typesetting
\usepackage{booktabs}       % professional-quality tables
\usepackage{amsfonts}       % blackboard math symbols
\usepackage{nicefrac}       % compact symbols for 1/2, etc.
\usepackage{lipsum}		% Can be removed after putting your text content
\usepackage{graphicx}
\usepackage{doi}

\usepackage{easyReview}
\usepackage{awesomebox}

\usepackage{graphicx}

\usepackage{breakurl}
\usepackage{xcolor}

\usepackage[protrusion=true, expansion=true, shrink=55, stretch=55, 
tracking=true, kerning=true, spacing=true, final]{microtype}

\usepackage[acronym]{glossaries}

\newacronym{HTTPS}{HTTPS}{HyperText Transfer Protocol Secure}
\newacronym{GDPR}{GDPR}{General Data Protection Regulation}
\newacronym{IT}{IT}{Information Technology}
\newacronym{MAC}{MAC}{Media Access Control}
\newacronym{HTML}{HTML}{HyperText Markup Language}
\newacronym{WASM}{WASM}{Web ASseMbly}
\newacronym{RSA}{RSA}{Rivest Shamir Adleman}
\newacronym{AESGCMSIV}{AES-GCM-SIV}{Advanced Encryption Standard Galois Counter Mode Synthetic Initialization Vector}
\newacronym{GCM}{GCM}{Galois Counter Mode}
\newacronym{PBKDF}{PBKDF}{Password-Based Key Derivation Function}
\newacronym{GPU}{GPU}{Graphics Processing Unit}
\newacronym{CSS}{CSS}{Cascading Style Sheets}
\newacronym{HIPAA}{HIPAA}{Health Insurance Portability and Accountability Act}

\usepackage{listings}

\definecolor{codegreen}{rgb}{0,0.6,0}
\definecolor{codegray}{rgb}{0.5,0.5,0.5}
\definecolor{codepurple}{rgb}{0.58,0,0.82}
\definecolor{codered}{rgb}{0.58,0,0}
\definecolor{backcolour}{rgb}{1, 1, 1}

\lstdefinestyle{jsstyle}{
    backgroundcolor=\color{backcolour},
    commentstyle=\color{codegreen},
    keywordstyle=\color{codered}\bfseries,
    numberstyle=\tiny\color{codegray},
    stringstyle=\color{codepurple},
    basicstyle=\ttfamily\footnotesize,
    breakatwhitespace=false,
    breaklines=true,
    captionpos=b,
    keepspaces=true,
    numbers=left,
    numbersep=5pt,
    showspaces=false,
    showstringspaces=false,
    showtabs=false,
    tabsize=2,
}
\lstdefinelanguage{JavaScript}{
  keywords={break, case, catch, continue, debugger, default, delete, do, else, finally, for, function, if, in, instanceof, let, new, return, switch, this, throw, try, typeof, var, void, while, with},
  otherkeywords={% Operators
    =, =>, ., \{, \}, [, ], ;
  },
  morecomment=[l]{//},
  morecomment=[s]{/*}{*/},
  morestring=[b]',
  morestring=[b]",
  sensitive=true,
  ndkeywords={class, export, boolean, throw, implements, import, this}
}

\newcommand{\wasm}{Web Assembly}
    
\title{Simple client-side encryption of personal information with \wasm{}%\thanks{Grants or other notes
}

\author{ \href{https://orcid.org/0000-0003-2642-519X}{\includegraphics[scale=0.06]{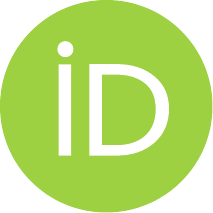}\hspace{1mm}Marco Falda} \\
    Department of Neuroscience\\
    University of Padova\\
    Padova, Italy \\
    \texttt{marco.falda@unipd.it} \\
    \And
    \href{https://orcid.org/0000-0002-8313-8384}{\includegraphics[scale=0.06]{orcid.pdf}\hspace{1mm}Angela Grassi} \\
    Veneto Institute of Oncology IOV–IRCCS\\
    Padova, Italy \\
    \texttt{angela.grassi@unipd.it} \\
}

\renewcommand{\shorttitle}{Simple client-side encryption of personal information with \wasm{}}

\hypersetup{
    pdftitle={Simple client-side encryption of personal information with WebAssembly},
    pdfauthor={Marco Falda, Angela Grassi},
    pdfkeywords={Client-side encryption, Rust language, End-to-end encryption},
}

\begin{document}

    \maketitle

    \begin{abstract}
        The HTTPS protocol has enforced a higher level of robustness to several attacks; however, it is not easy to set up the required certificates on intranets, nor is it effective in the case the server confidentiality is not reliable, as in the case of cloud services, or it could be compromised.
        
        A simple method is proposed to encrypt the data on the client side, using \wasm{}. It never transfers data to the server as clear text. Searching fields in the server is made possible by an encoding scheme that ensures a stable prefix correspondence between ciphertext and plaintext.
        
        The method has been developed for a semantic medical database, and allows accessing personal data using an additional password while maintaining non-sensitive information in clear form. \wasm{} has been chosen to guarantee the fast and efficient execution of encrypting/decrypting operations and because of its characteristic of producing modules that are very robust against reverse engineering.
        
        The code is available at \href{https://github.com/mfalda/client-encdec}{https://github.com/mfalda/client-encdec}.
    \end{abstract}
    % M: applicati tutti i suggerimenti nell'abstract
    
    
    \section{Introduction}
    \label{sec:intro}

    %\notebox{
    %    \textbf{EasyReview:}
    %    \begin{description}
    %        \item[Add] \add{adding a text}.
    %        \item[Remove] \remove{removing a text}.
    %        \item[Replace] \replace{replacing a text}{with another}.
    %        \item[Notes] \alert{a note}.
    %        \item[Highlight] \highlight{highlighting a text}
    %        \item[Comments] \comment{annotating}{this text}.
    %    \end{description}
    %}
    
    % Introduce the scientific background and the motivation for developing the software.
    Cloud storage is one of the most popular forms of data storage, with an ever-increasing impact on society, businesses, and organizations. In 2020, Cybersecurity Ventures predicted more than 200 zettabytes ($200 \times 2^{70}$ bytes) of digital information on and off the Web by 2025 \cite{Zettabytes2020}. However, it is not clear how these data are managed and protected. Several large incidents have already been reported \cite{CyberTalk2022}. In addition, many more pass under silence or are not detected.
    
    For this reason, there is an increasing interest in end-to-end encryption \cite{Bai2020} and also zero-knowledge encryption, in which user data are encrypted with an asymmetric algorithm such as \gls{RSA}. Common implementations use a public key to encrypt (chunks of) data on the client before they are uploaded to the server; in downloads, data are decrypted on the client with the corresponding private key. The private key is never transmitted to the server. A major disadvantage of this technique is that in the case the user loses his/her password, all data are lost.
    
    Besides the specific file storage applications, the possibility to deal with the personal data contained in files in a secure way is fundamental for online databases that are designed not only to allow statistical analysis of aggregate data but also for the management of people. In particular, health data management systems store sensitive data, including medical records, and it is necessary for medical personnel to be able to clearly identify a patient, for reference, and to update information. Medical data leaks can have serious consequences. For example, in 2018, 1.5 million SingHealth patients’ non-medical personal data were taken from the country’s healthcare system \cite{Hambleton2018}.
    
    Recently, in the European Union, a new regulation, \gls{GDPR} \cite{EUdataregulations2018}, has been imposed to protect the privacy of citizens. It mandates the careful management of sensitive data and the so-called right to erasure (``right to be forgotten'', art. 17). Therefore, healthcare data must be protected and personal information must be hidden from cloud service providers \cite{Huang2019,Abbas2014}. To deal with data protection, \gls{GDPR} suggests the use of encryption, and it is not mandatory to publicly report data leaks of (adequately) encrypted data.
    
    %% Explain why the software is important, and describe the exact (scientific) problem(s) it solves.
    Several Italian University Hospital Departments and Institutes for Research Hospitalization and Healthcare (IRCCS) use intranets but do not provide a Certification Authority to allow setting up the \gls{HTTPS} protocol for secure communications. Intranets are usually protected by multiple firewalls and other cyber-security measures \cite{Cabral2019} and devices are subject to \gls{MAC} address registration to be able to connect to the network; however, it may be possible for an insider to intercept communications by \gls{MAC} spoofing anyway.
    
    For this reason, we propose adding a new protection layer to the usual authentication method based on username and password by encrypting sensible data right into the client before transmitting them to the server. Another advantage is the possibility of strengthening the protection of sensitive information, access to which is reserved only for certain users.
    
    % Indicate in what way the software has contributed (or how it will contribute in the future) to the process of scientific discovery; if available, this is to be supported by citing a research paper using the software.
    There are only a few examples of client-side encryption for Web pages. In \cite{Morse2011} plaintexts are encrypted before being sent to the server, and a hash on the server is used to verify the correctness of the user password; however, it is not clear how encrypted data are searched in the server or if it is possible at all. Another application to provide end-to-end encryption is Mylar \cite{Mylar2014}, which allows searching for keywords over the encrypted document on the server. Its drawbacks are a more complex architecture and the necessity to install an ad hoc plug-in and to adopt a specific web framework.
    
    % Gap and proposal
    In this paper, we propose to encrypt sensitive data using \wasm{} modules, which are much more efficient with respect to JavaScript. Moreover, we use an encryption scheme that allows searching encrypted fields with a limited amount of ambiguity in the matches.
    
    %% Provide a description of the experimental setting (how does the user use the software?).
    
    %% Introduce related work in literature (cite or list algorithms used, other software etc.).

    \section{Software description}
    \label{sec:descr}
    
    \subsection{Software Architecture}
    \label{sec:arch}
    
    %Give a short overview of the overall software architecture; provide a pictorial component overview or similar (if possible). If necessary provide implementation details.
    
\begin{figure*}
        \centering
        \includegraphics[width=12cm]{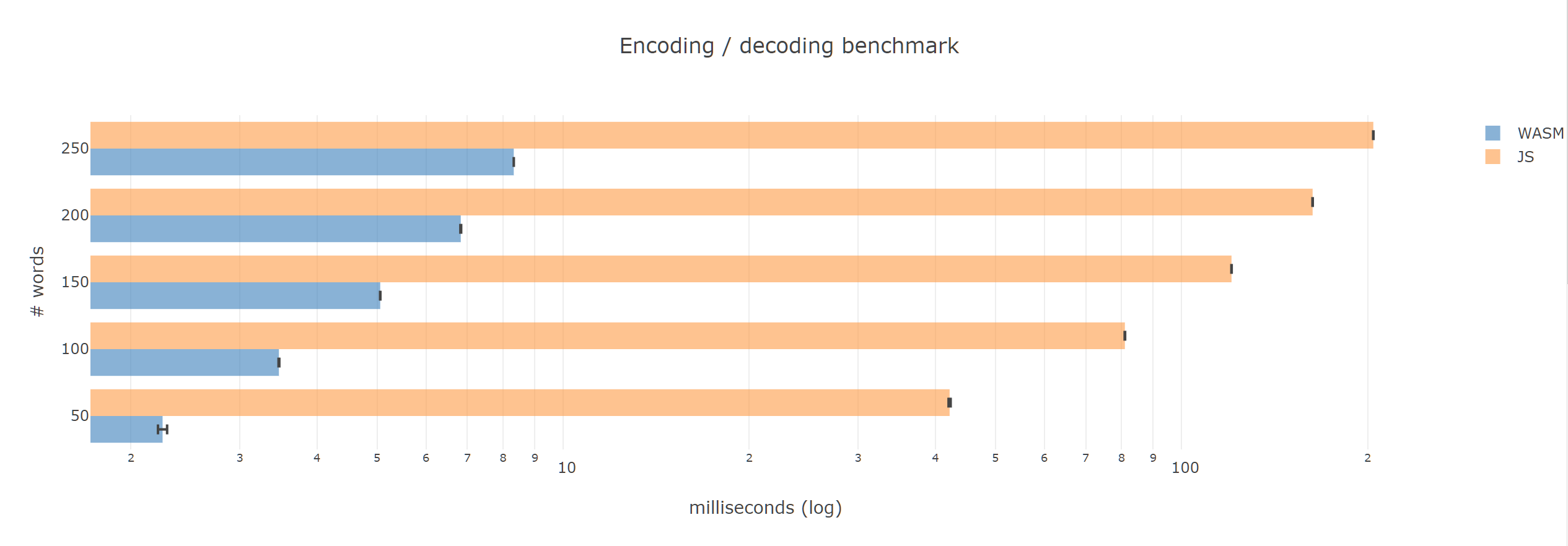}
        \caption{Comparison between native JavaScript and \wasm{} implementation.}
        \label{Fig:Speed}
    \end{figure*}

    %Describe the software in as much as is necessary to establish a vocabulary needed to explain its impact.
    The library provides two functions for encrypting and decrypting \gls{HTML} texts; they are simple enough to allow developers with minimal \gls{HTML} knowledge to integrate them into their Web forms and pages. The library has been developed using \wasm{}, a new language that allows executing binary code in client browsers; the resulting code is fast, as shown in Figure~\ref{Fig:Speed}.
    
\begin{figure}
        \centering
        \includegraphics[width=5cm]{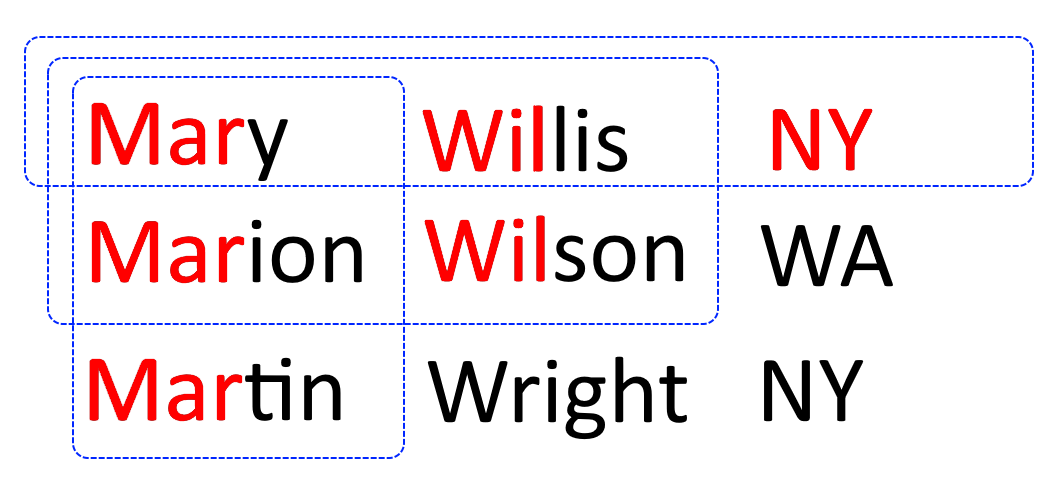}
        \caption{Reduction of the results sets by adding search terms prefixes.}
        \label{Fig:pref-search}
    \end{figure}

\begin{figure*}
        \centering
        \includegraphics[width=0.75\textwidth]{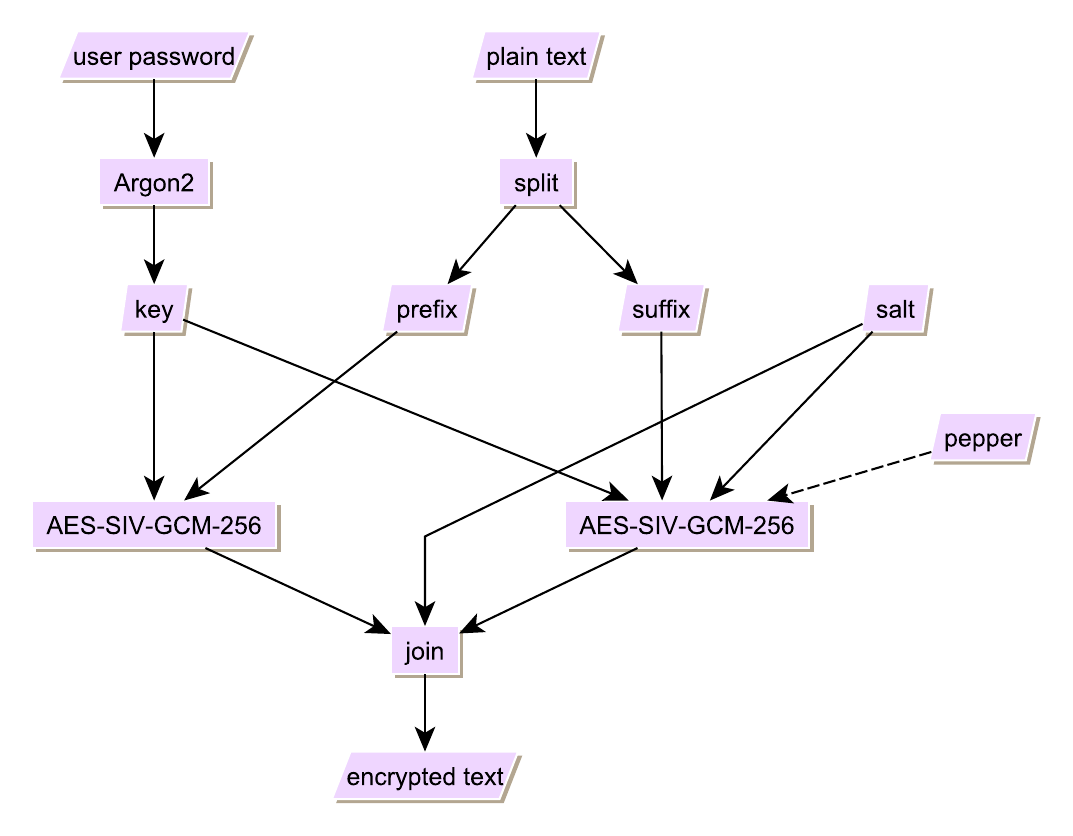}
        \caption{Encryption schema.}
        \label{Fig:Enc_schema}
    \end{figure*}

    The binary code can be produced by several programming languages: C/C++, C\#, Go, and Rust. Among them, Rust is one of the most robust and performant programming languages, thanks to its native memory allocation management and the borrow-checker facility, which denies unsafe code \cite{Rust2010}. To allow searching for encrypted fields directly on the server, plaintexts are partitioned into a prefix and a suffix (more than two substrings could be created and then shuffled together in a predetermined way). In this work, we propose to encrypt the prefix with a deterministic algorithm and to apply a salt and possibly a ``pepper'' \cite{Manber96} to the suffix, to improve randomness against cryptanalyses (see Figure~\ref{Fig:Enc_schema}). In this way, it is possible to perform prefix searches on the server and obtain a limited result set, which can be refined by adding other fields. The same prefix ambiguity ensures that even in the case that the prefixes are decrypted, their shortness will not produce meaningful data (see Figure~\ref{Fig:pref-search}).

    Regarding encryption algorithms, the \gls{AESGCMSIV} algorithm with a 256-bit key has been chosen, as it seems to have reasonable robustness \cite{Bose2018,Secara2020}.
    A 256 bit key corresponds to a 32-character password, which is too long for users; moreover, it could be too simple, that is, it could have a little amount of entropy and, therefore, be too easy to discover. Therefore, the password is passed to a \gls{PBKDF} function in order to generate a 32-character key from it; we chose the Argon2(id) algorithm, one of the most robust against both \gls{GPU} cracking and side-channel attacks \cite{argon2}.
    
    Since all data are encrypted with the same key, it is important to ensure its correctness. In literature, this problem is solved, for example, by storing a hash in the server and then by checking it on the client \cite{Morse2011}. Another possibility, adopted in our case, is to store three or more encrypted check-words in the server and then decrypt and verify them in the client; these check-words may have a prefix of length zero.
    
    The initial encryption of the entries can be performed in Node.js, using the same \wasm{} module. Moreover, since the Rust implementation is very efficient, it is possible to insert a complete encryption/decryption/validation cycle for each entry.
    
    \subsection{Software Functionalities}
    \label{sec:fun}
    % Present the major functionalities of the software.
    The library has two (main) functions: \textbf{encryptText} and \textbf{decryptText}. Both accept three parameters:
    \begin{description}
        \item[password] the user password;
        \item[text] the plaintext to be encrypted or the ciphertext to be decrypted;
        \item[pref\_length] the prefix length (0 for a non-deterministic algorithm).
    \end{description}
    
    Since the user password must not be passed to the server, it must be stored locally, for example, in the \gls{HTML}5 sessionStore facility.

    \section{Illustrative Examples}
    \label{sec:examples}
    
    % Provide at least one illustrative example to demonstrate the major functions.
    The functions are very simple and can be used in \gls{HTML} events such as \texttt{onsubmit}, \texttt{onfocus} or \texttt{onblur}. For example, if there is a form with text input marked with the \gls{CSS} class ``encrypted'', the functions could be associated in this way:
    
    \begin{lstlisting}[language=JavaScript]
        let ctrls = document.getElementsByClassName("encrypted");
        [].forEach.call(ctrls, (el) => {
            el.onfocus = (event) => {
                document.getElementById("firstname").value = decryptText("O&3p5#2", event.target.value);
            });
            el.onblur = (event) => {
                document.getElementById("firstname").value = encryptText("O&3p5#2", event.target.value);
            });
        });
    \end{lstlisting}
    
    If we want to decrypt the existing text that has been retrieved encrypted from the server, the \texttt{onready} event could be exploited, provided that it has been wrapped in \texttt{<span> } tags marked with the \gls{CSS} class ``encrypted'':
    
    \begin{lstlisting}[language=JavaScript]
        document.addEventListener("DOMContentLoaded", () => {
            let encTexts = document.querySelectorAll("span.encrypted");
            [].forEach.call(encTexts, decryptText("O&3p5#2", event.target.value));
        });
    \end{lstlisting}

    \section{Impact}
    \label{sec:impact}
    
    %\textbf{This is the main section of the article and the reviewers weight the description here appropriately}: sistemare meglio in base ai suggerimenti della rivista.
    
    %Indicate in what way new research questions can be pursued as a result of the software (if any).
    
    %Indicate in what way, and to what extent, the pursuit of existing research questions is improved (if so).
    
    %Indicate in what way the software has changed the daily practice of its users (if so).
    
    %Indicate how widespread the use of the software is within and outside the intended user group.
    
    %Indicate in what way the software is used in commercial settings and/or how it led to the creation of spin-off companies (if so).

    We needed a method to encrypt data on the client and to have the communication automatically protected. Encrypting sensitive data on the client should mitigate ``man-in-the-middle'' attacks and should protect personal data even in the case the server is compromised \cite{Mylar2014,Contini2015}.
    
    The possibility of searching for sensible data in a database on the server while keeping them protected is an important aspect. This was possible by designing an encoding scheme that maintained determinism in the prefix while adding a salt in the suffix. In this way, even if there are ambiguities, they can be reduced as long as search terms are added (see Figure~\ref{Fig:pref-search}).
    
    As in the case of the cited Morse proposal \cite{Morse2011}, this is an additional protection layer with respect to the common username and password authentication managed by the web server or the web platform. An additional two-factor authentication should further enforce this security layer.
    
    The library could also be extended to deal with dates, locations, and other sensible information. For example, in \gls{HIPAA} \cite{HIPAA97} admission dates and other timestamps are considered to belong to the limited data set; therefore, they should also be protected. Similarly, geographical locations could also be considered worth protecting. Two approaches could be devised: aggregate points into a higher entity or apply a constant offset. The aim would be to be able to keep such data clear in order to perform statistical analyses on them. The first approach seems to be more appropriate for geographical coordinates, for example, using isopopulated districts, while the latter is more appropriate for dates. This is evident from the reasoning that points that have been uniformly translated on a map could be easily repositioned in the correct origin considering the fact that clusters of points correspond to the towns. For dates, the considerations are more subtle: Sometimes even a few years or months could make a difference in human evolution, and, reasoning about the same patient, it is better to maintain time distances.

    \section{Conclusion}
    \label{sec:end}
    
    % 5a. restate problem: 2 s.
    In this work, a new method for encrypting sensitive data has been proposed. The forms compiled in the client are encrypted using a \wasm{} module and therefore the server cannot store plain text. In this way, data are protected during transmission and can also be securely stored on untrusted servers such as cloud services.
    
    % 5b. interesting part of the work: 2 s.
    Moreover, the proposed encoding allows the retrieval of data from the server. Encrypting functions have been developed in Rust and therefore are much faster and efficient with respect to their equivalent JavaScript version. They are also more resistant to decompilation.
    
    % 5c. future development: 3 s.
    The next step could be protection of \wasm{} modules from ``black-box" usage, that is, the exploitation of the \wasm{} function to generate encrypted texts in order to perform a cryptanalysis, even if it would be limited to short prefixes. Another enhancement could be the application of encryption methods that would allow statistical computations directly on the cyphertexts \cite{Homomorphic2018}.
    
    \bibliographystyle{spphys}
    \bibliography{bib}   % name your BibTeX data base

\begin{thebibliography}{10}
\providecommand{\url}[1]{{#1}}
\providecommand{\urlprefix}{URL }
\expandafter\ifx\csname urlstyle\endcsname\relax
  \providecommand{\doi}[1]{DOI \discretionary{}{}{}#1}\else
  \providecommand{\doi}{DOI \discretionary{}{}{}\begingroup \urlstyle{rm}\Url}\fi

\bibitem{Zettabytes2020}
S.~Morgan.
\newblock {The World Will Store 200 Zettabytes Of Data By 2025}.
\newblock https://cybersecurityventures.com/the-world-will-store-200-zettabytes-of-data-by-2025/ (2020)

\bibitem{CyberTalk2022}
{CyberTalk.org}.
\newblock Top 5 cloud security breaches (and lessons).
\newblock https://www.cybertalk.org/2022/04/26/top-5-cloud-security-breaches-and-lessons/ (2022)

\bibitem{Bai2020}
W.~Bai, M.~Pearson, P.G. Kelley, M.L. Mazurek, in \emph{2020 IEEE European Symposium on Security and Privacy Workshops (EuroS\&PW)} (2020), pp. 210--219.
\newblock \doi{10.1109/EuroSPW51379.2020.00036}

\bibitem{Hambleton2018}
H.~Steve, A glimpse of 21st century care, Australian Journal for General Practitioners \textbf{47}(10), 670 (2018)

\bibitem{EUdataregulations2018}
2018 reform of eu data protection rules (2018)

\bibitem{Huang2019}
Z.~Huang, J.~Lai, W.~Chen, T.~Li, Y.~Xiang, Data security against receiver corruptions: Soa security for receivers from simulatable dems, Information Sciences \textbf{471}, 201 (2019).
\newblock \doi{https://doi.org/10.1016/j.ins.2018.08.059}

\bibitem{Abbas2014}
A.~Abbas, S.U. Khan, A review on the state-of-the-art privacy-preserving approaches in the e-health clouds, IEEE Journal of Biomedical and Health Informatics \textbf{18}(4), 1431 (2014).
\newblock \doi{10.1109/JBHI.2014.2300846}

\bibitem{Cabral2019}
W.~Cabral, C.~Valli, L.~Sikos, S.~Wakeling, in \emph{2019 International Conference on Computational Science and Computational Intelligence (CSCI)} (2019), pp. 166--171.
\newblock \doi{10.1109/CSCI49370.2019.00035}

\bibitem{Morse2011}
R.E. Morse, P.~Nadkarni, D.A. Schoenfeld, et~al., Web-browser encryption of personal health information, BMC Med Inform Decis Mak \textbf{11}(70) (2011)

\bibitem{Mylar2014}
R.A. Popa, E.~Stark, S.~Valdez, J.~Helfer, N.~Zeldovich, H.~Balakrishnan, in \emph{11th USENIX Symposium on Networked Systems Design and Implementation (NSDI 14)} (USENIX Association, Seattle, WA, 2014), pp. 157--172

\bibitem{Rust2010}
G.~Hoare.
\newblock Project servo.
\newblock Mozilla Annual Summit (2010)

\bibitem{Manber96}
U.~Manber, A simple scheme to make passwords based on one-way functions much harder to crack, Computers \& Security \textbf{2}(15), 171 (1996).
\newblock \doi{10.1016/0167-4048(96)00003-x}

\bibitem{Bose2018}
P.~Bose, V.T. Hoang, S.~Tessaro, in \emph{Advances in Cryptology -- EUROCRYPT 2018}, ed. by J.B. Nielsen, V.~Rijmen (Springer International Publishing, Cham, 2018), pp. 468--499

\bibitem{Secara2020}
I.A. Secara, Zoombombing - the end-to-end fallacy, Network Security \textbf{2020}(8), 13 (2020).
\newblock \doi{https://doi.org/10.1016/S1353-4858(20)30094-5}

\bibitem{argon2}
J.~Wetzels.
\newblock Open sesame: The password hashing competition and argon2 (2016).
\newblock \doi{10.48550/ARXIV.1602.03097}.
\newblock \urlprefix\url{https://arxiv.org/abs/1602.03097}

\bibitem{Contini2015}
S.~Contini.
\newblock Method to protect passwords in databases for web applications.
\newblock Cryptology ePrint Archive, Paper 2015/387 (2015).
\newblock \urlprefix\url{https://eprint.iacr.org/2015/387}.
\newblock \url{https://eprint.iacr.org/2015/387}

\bibitem{HIPAA97}
B.K. Atchinson, D.M. Fox, {The Politics Of The Health Insurance Portability And Accountability Act}, Health Affairs \textbf{3}(16), 146 (1997).
\newblock \doi{10.1377/hlthaff.16.3.146}

\bibitem{Homomorphic2018}
K.~Gai, M.~Qiu, Blend arithmetic operations on tensor-based fully homomorphic encryption over real numbers, IEEE Transactions on Industrial Informatics \textbf{14}(8), 3590 (2018).
\newblock \doi{10.1109/TII.2017.2780885}

\end{thebibliography}
    
\end{document}